\documentclass[aps,prl,reprint,superscriptaddress]{revtex4-1}
\usepackage{graphicx}
\usepackage{color}
\usepackage{xcolor}
\usepackage{bm}
\begin{document}

\title{Flow-Arrest Transitions in Frictional Granular Matter}

\author{Ishan Srivastava}
\affiliation{Sandia National Laboratories, Albuquerque, New Mexico 87185, USA}
\author{Leonardo E. Silbert}
\affiliation{School of Math, Science, and Engineering, Central New Mexico Community College, Albuquerque, New Mexico 87106, USA}
  \author{Gary S. Grest}
\author{Jeremy B. Lechman}
\affiliation{Sandia National Laboratories, Albuquerque, New Mexico 87185, USA}

\begin{abstract}

The transition between shear-flowing and shear-arrested states of frictional granular matter is studied using constant-stress discrete element simulations. By subjecting a dilute system of frictional grains to a constant external shear stress and pressure, friction-dependent critical shear stress and density are clearly identified with both exhibiting a crossover between low and high friction. The critical shear stress bifurcates two nonequilibrium steady states:~(i) steady state shear flow characterized by a constant deformation rate, and (ii) shear arrest characterized by temporally decaying creep to a statically stable state. The onset of arrest below critical shear stress occurs at a time $t_{c}$ that exhibits a heavy-tailed distribution, whose mean and variance diverge as a power law at the critical shear stress with a friction-dependent exponent that also exhibits a crossover between low and high friction. These observations indicate that granular arrest near critical shear stress is highly unpredictable and is strongly influenced by interparticle friction. 
\end{abstract} 

\maketitle

Granular materials exhibit complex deformation and rheological phenomena due to the discrete character of their constituent particles and dissipative frictional interactions between them. It exists---or can co-exist---in fluidlike and solidlike states, and transitions between these states (i.e., yielding or arrest) are often induced by applying normal and deviatoric stresses at the boundary. Historically, the yielding of granular matter has been modeled using failure criteria such as the Mohr-Coulomb theory, which predicts yield when the ratio of deviatoric to normal stress exceeds a material-dependent threshold. Later, plasticity theories such as critical state soil mechanics~\cite{schofield1968critical} were developed that define a \emph{critical} state  at which granular matter yields at a constant critical shear stress, pressure and volume fraction. However, these theories are restricted to quasistatic rate-independent yielding of granular matter. Recent advances have addressed rate dependence of steady granular flows by extending such plasticity theories to viscoplastic rheology~\cite{Jop:2006gh}. These rheological models have been successful in describing yielding and steady dense granular flows in various geometries~\cite{GDRMiDi:2004jd}. 

However, transient granular dynamics during the initiation~\cite{CourrechduPont:2005id,Jop:2007ih,Andrade:2012iq} or arrest of a steady flow~\cite{Silbert:2003gk,Ciamarra:2011ju,Pastore:2011dd} exhibit remarkable features that are unexplained by current models, such as anomalous velocity profiles~\cite{Silbert:2003gk,CourrechduPont:2005id}, hysteresis between flow initiation and flow arrest~\cite{Silbert:2003gk}, transient dilatancy~\cite{Andrade:2012iq}, and a breakdown of isostaticity~\cite{Kruyt:2010cq}. In addition, granular matter often exhibits slow unsteady creep flows below a critical stress~\cite{Nguyen:2011hm,Amon:2012ip,Ishan.2017}. These phenomena are not captured by existing theories. 

The mechanics of granular matter near the static-dynamic transition (i.e., quasistatic flows) are also highly intermittent and stochastic, and remain unexplained by deterministic theories described previously. Growing avalanches~\cite{LeBouil:2014ec,Lin:2015gn}, localized plastic rearrangements~\cite{Amon:2012ip}, and fluctuations in stresses~\cite{Dalton:2005jk} dominate the transition between flowing and arrested states. Although mean-field models have been developed to predict the statistics of such phenomena~\cite{Dahmen:2011co}, their applicability in continuum granular rheology, especially during transient dynamical evolution, remains unclear. 

It is, therefore, critical that existing theories be extended to account for the transient dynamics near the granular flow-arrest transition, while also addressing the highly stochastic nature of flow and deformation in that regime. In this Letter, we highlight the transient stochastic dynamics of frictional granular matter near the flow-arrest transition by approaching a critical shear stress from a flowing granular state. The transition from a flowing state to an arrested state is characterized by diverging times that are highly dependent on the friction between particles. Although this flow-arrest transition always occurs at the same friction-dependent density and shear stress, the times to arrest below critical shear are widely distributed near the transition even for similar systems, indicating the need for additional microscopic details beyond the bulk state of stress to characterize the statistical nature of the transition.

We probe the dynamics of the shear-induced flow-arrest transition through simulations that impose pressure and shear stress on a bulk frictional granular system, and allow dynamical evolution to a flowing or an arrested state depending upon the applied stress. In this way, the critical stress is approached along paths of constant applied pressure and shear stress, while simultaneously allowing the density to evolve to a critical state via dilation or compaction. Although several stress-controlled studies~\cite{DaCruz:2002im,Dijksman:2011is,Nguyen:2011hm,Boyer:2011hf,Amon:2012ip} have reported incipient granular yield and flow with remarkable results, complications such as the presence of gravity, shear banding~\cite{Schall:2010fu} and boundary effects~\cite{Silbert:2002ii,Jop:2005ee,Depken:2007cq} make the mechanical analysis challenging. The present simulations, in contrast, are able to capture the bulk behavior of the granular matter near critical stress. 

\begin{figure}[b!]
\includegraphics[width=3.0in]{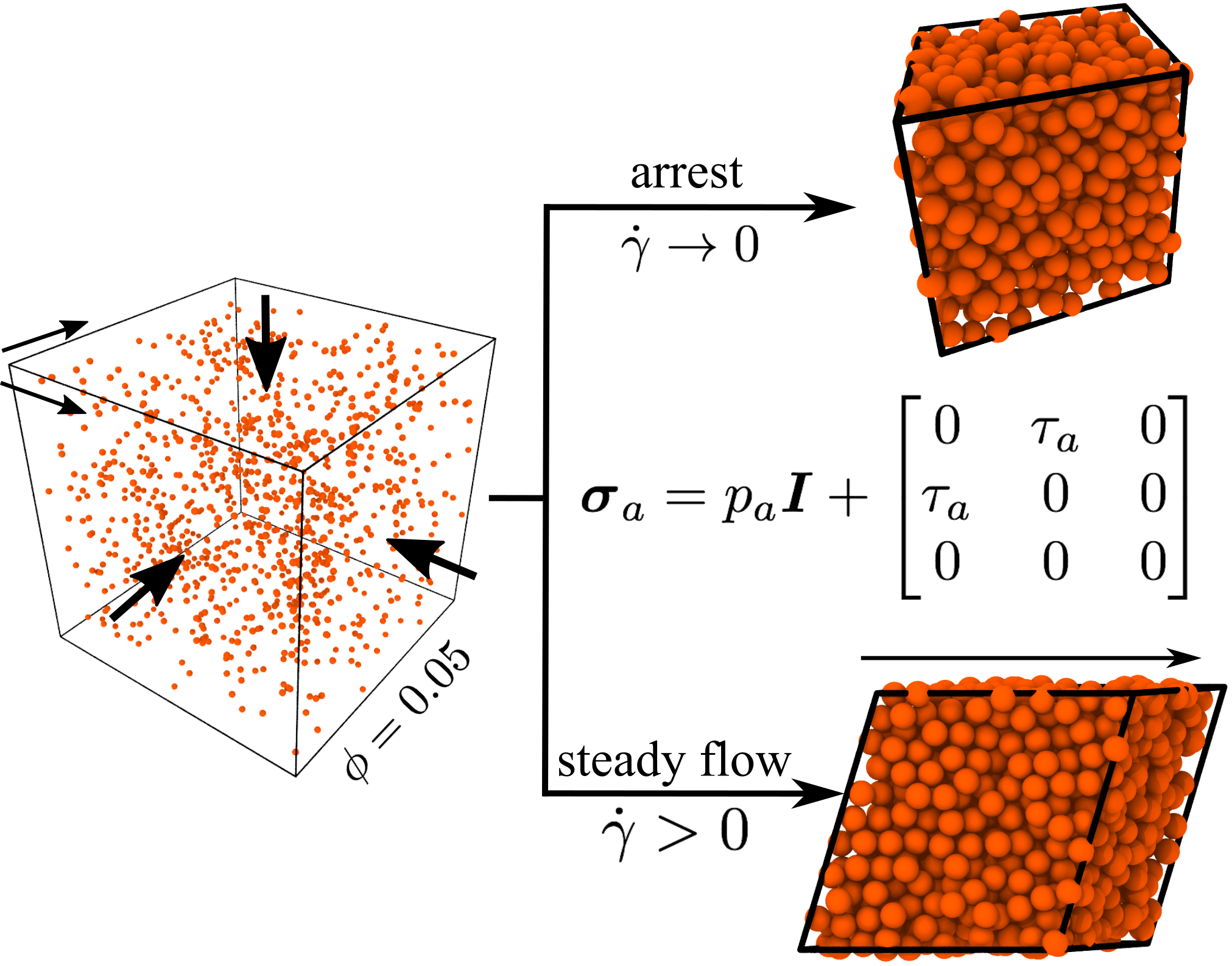}
\caption{Schematic of the simulation method. The left image represents a starting low density configuration at $\phi=0.05$. The arrows around the black periodic box represent applied stress tensor $\boldsymbol{\sigma}_{a}$, which is composed of pressure $p_{a}$ and shear $\tau_{a}$ components. The two images on the right represent two possible long-term dynamics: (top) shear arrest characterized by a vanishing strain rate $\dot{\gamma}\to0$, and (bottom) steady flow characterized by a steady nonzero strain rate.} 
\label{fig1} 
\end{figure}

\emph{Numerical setup.---} We subject dilute granular systems, prepared at an initial density $\phi\!=\!0.05$, to prescribed values of applied pressure $p_{a}$ and shear $\tau_{a}$ at zero gravity, such that the applied stress tensor $\boldsymbol{\sigma}_{a}$ is constrained only by (i) $(1/3)\!\sum\sigma_{a,ii}\!=\!p_{a}$, (ii) $\sigma_{a,ij}\!=\!\tau_{a}$ for $i,j\!=\!1,\!2$ and $2,\!1$, and (iii) $\sigma_{a,ij}\!=\!0$ for all other indices $i\!\neq\!j$. The pressure is fixed at $p_{a}\!=\!10^{-4}k_{n}/d$ to simulate the asymptotic rigid particle regime, where $k_{n}$ is the Hookean spring constant between contacting particles and $d$ is the average particle diameter. The applied shear $\tau_{a}$ is modulated such that two distinct dynamical responses at long times are observed: (i) arrested dynamics characterized by a vanishing bulk strain rate $\dot{\gamma}\to0$, and (ii) steady flow characterized by a constant, nonzero strain rate. The simulation method is described schematically in Fig.~\ref{fig1}. Here $\dot{\gamma}$ is the second invariant of a dynamically evolving bulk 3D strain rate tensor $\dot{\gamma}_{ij}$.  

 A constant stress tensor is applied and maintained at the system boundaries using the dynamical equations of motion of Parinello-Rahman (PR)~\cite{Parrinello:1981it}, which treats the three basis vectors of a periodic cell as dynamical variables. In particular, the PR method controls the second Piola-Kirchoff stress (which is the thermodynamic conjugate of the Lagrangian strain~\cite{Souza:1997fj}), but not the Cauchy stress $\sigma_{ij}$, which is defined in the deformed state~\footnote{Piola-Kirchoff stress $\bm{\sigma}_{a}$ is related to Cauchy stress $\bm{\sigma}$ via deformation gradient $\bm{F}$ as: $\bm{\sigma}\!=\!(1/J)\bm{F}\bm{\sigma}_{a}\bm{F}^{T}$, where Jacobian $J=\text{det}\left[\bm{F}\right]$.}. However, Cauchy stress is an operationally meaningful measure of stress typically measured in experiments, and is calculated as the sum of internal contact and kinetic stresses in granular systems. The dynamics of the system are driven by the imbalance between applied and measured stress, and mechanical equilibrium is achieved upon their balance.  The two invariants of Cauchy stress, (i) pressure $p\!=\!(1/3)\!\sum\sigma_{ii}$ and (ii) shear $\tau\!=\!\left(0.5\sum_{i,j}\tau_{ij}\tau_{ij}\right)^{1/2}$, define the stress ratio $\mu\!=\!\tau/p$. Here $\tau_{ij}\!=\!\sigma_{ij}\!-\!p\delta_{ij}$ is the deviatoric stress. Such stress-based methods have been previously used to study collective jamming~\cite{Smith:2014co} and creep~\cite{Ishan.2017} in granular matter. 
 
 \begin{figure}[b!]
\includegraphics[width=3.0in]{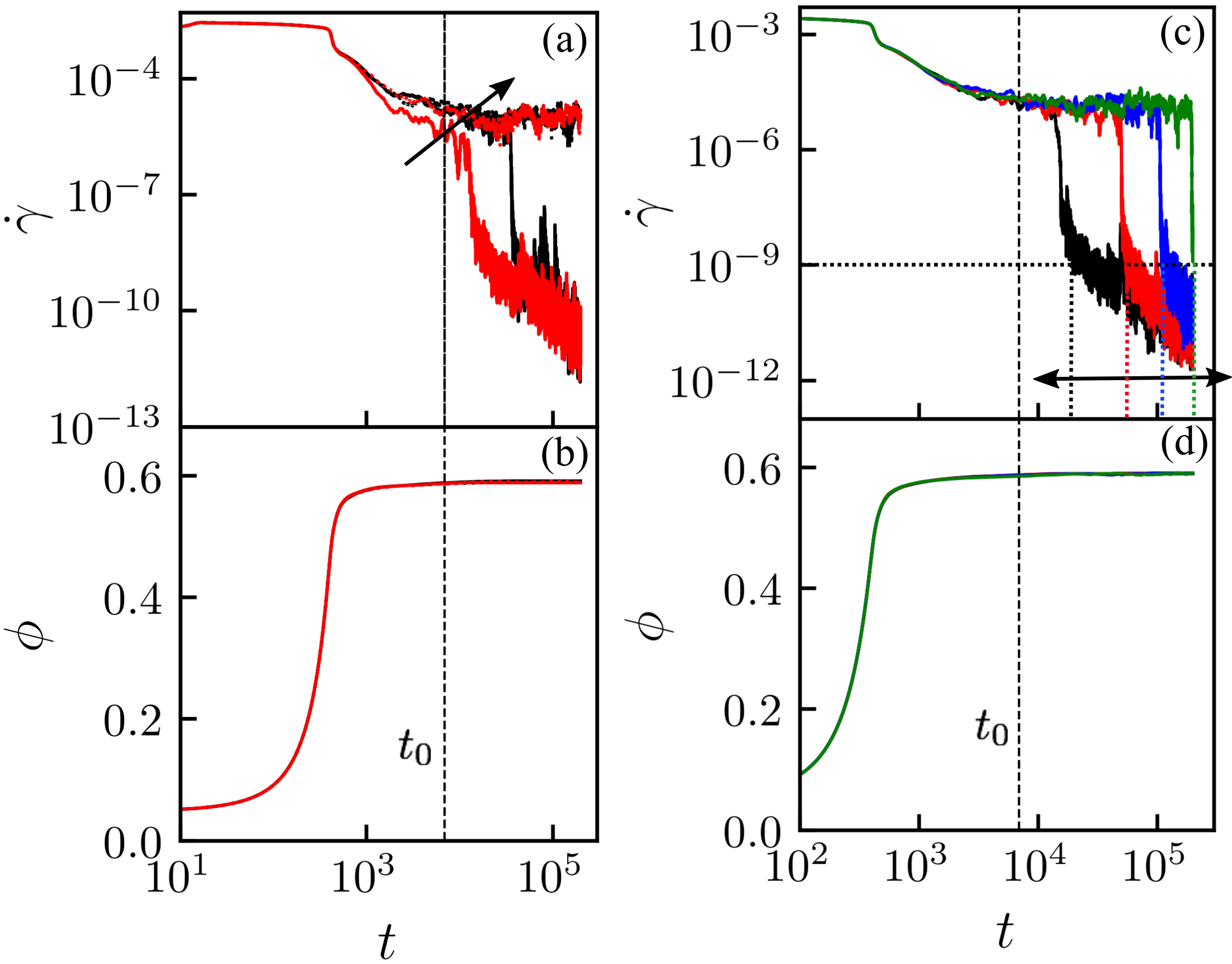}
\caption{Variation of (a) $\dot{\gamma}$ and (b) $\phi$ with time $t$ for the same starting state and $\mu_{s}\!=\!1.0$, subjected to varying $\tau_{a}$. The arrow in (a) represents the direction of increasing $\tau_{a}$. The top two dashed curves represent steady flowing states, whereas the bottom two solid curves represent arrested dynamics at long times. (c) and (d): Four curves illustrate arrested dynamics for the same $\tau_{a}$ applied to four different starting states and $\mu_{s}\!=\!1.0$. The vertical dashed lines in (a) - (d) represent $t_{0}$ defined in Eq.(~\ref{eq1}). The horizontal dotted line in (c) represents the criteria for determining onset of arrest $t_{c}$, whereas the vertical dashed lines denote $t_{c}$ for each of the four cases.} 
\label{fig2} 
\end{figure}
 
The simulations consist of $10^{4}$ dispersed spherical particles, whose sizes are uniformly distributed between $0.9d$ and $1.1d$. The particles interact via a linear spring-dashpot contact mechanical model with elastic and viscous normal and tangential forces~\cite{Silbert.2001}. A Coulomb friction model is used to calculate tangential frictional forces between two contacting particles, and the coefficient of friction $\mu_{s}$ was varied between $0.001$ and $1.0$ to study the effect of friction on the flow-arrest transition. Time is normalized by the characteristic timescale $\sqrt{m/k_n}$, where $m$ is the average mass of a particle, and the simulation time step is set to $0.02\sqrt{m/k_n}$. The Hookean spring constant $k_{n}$ sets the scale for energy and stress; therefore, all stresses are scaled by $k_{n}/d$.

\emph{Dynamical evolution.---}Figure~\ref{fig2}(a) demonstrates the typical evolution of $\dot{\gamma}$ with time $t$ for the same starting state, but varying applied shear. At early times the dilute system experiences large strain as it is unable to resist the applied load. As time progresses and the system compacts under pressure [see Fig.~\ref{fig2}(b)], two distinct dynamical scenarios emerge: (i) for high $\tau_{a}$ the system continues to flow indefinitely at a constant strain rate; (ii) for low $\tau_{a}$ the strain rate abruptly drops several orders of magnitude as the system enters a dynamically arrested regime characterized by creep to a final mechanically stable state. Interestingly, the bifurcation in strain rate evolution based on the magnitude of applied shear is not reflected in the density evolution, as can be seen by overlapping curves in Fig.~\ref{fig2}(b). This demonstrates that density depends only on applied pressure, which is the same across all simulations, and shear stress has a negligible effect on its evolution for the weakly polydisperse spheres simulated here. 

 \begin{figure}[b!]
\includegraphics[width=3.0in]{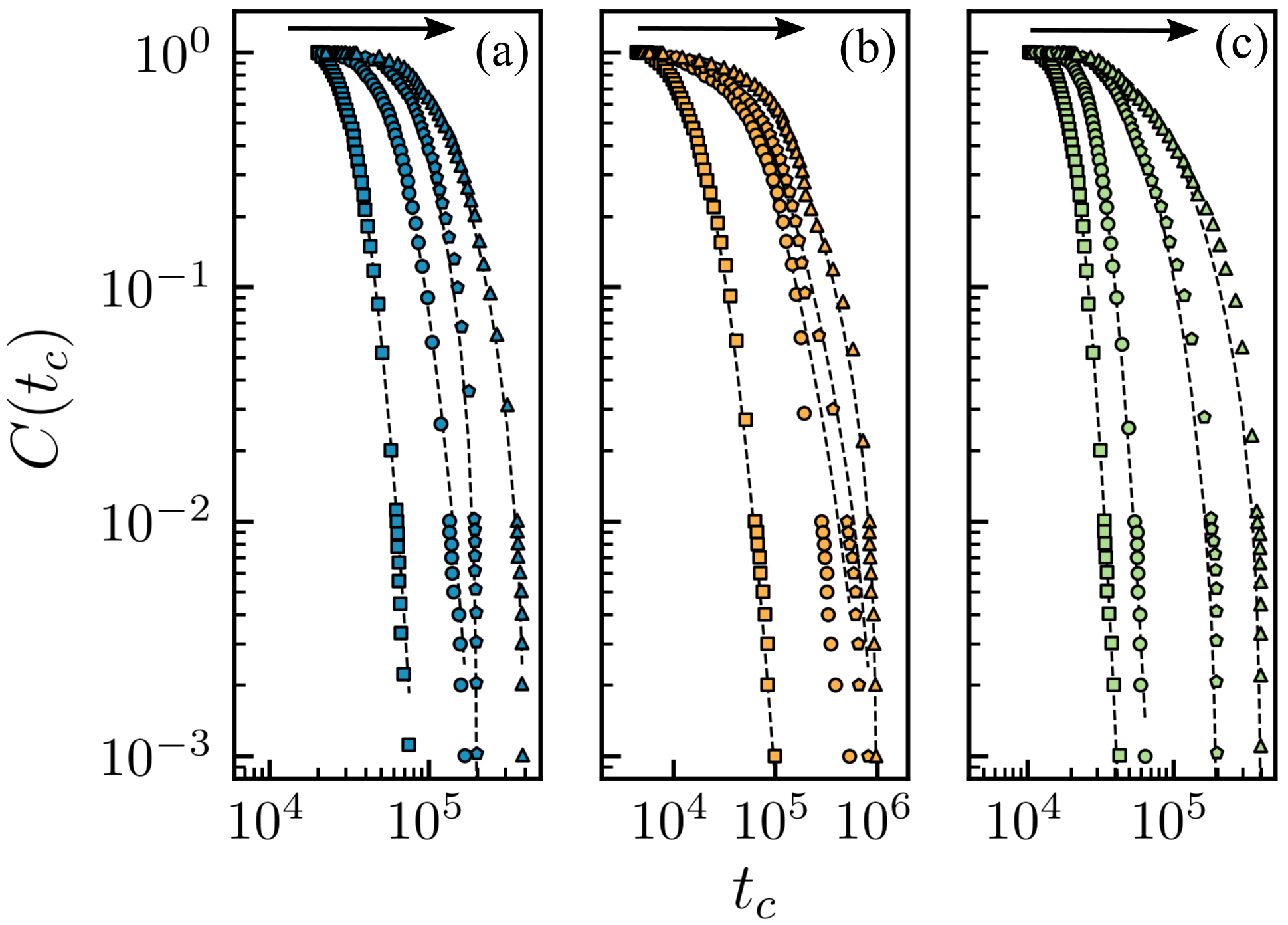}
\caption{Complementary cumulative distribution of arrest times $C(t_c)$ for friction (a) $\mu_{s}\!=\!10^{-3}$, (b) $\mu_{s}\!=\!6.5\!\times\!10^{-2}$, and (c) $\mu_{s}\!=\!1.0$ at different stress ratios $\mu$. The arrow represents the direction of increasing $\mu$, and its values from left to right are (a) $\left(0.082,0.109,0.116,0.119\right)$, (b) $\left(0.243,0.246,0.247,0.248\right)$, and (c) $\left(0.327,0.339,0.348,0.35\right)$. The dashed lines represent the best estimate fit for a log-normal distribution.}
\label{fig3} 
\end{figure}

\emph{Statistics of arrest.---}Although distinct regimes of granular flow or arrest emerge as a function of applied shear, significant statistical variations accompany the onset of arrest. In Fig.~\ref{fig2}(c) the variation of $\dot{\gamma}$ with $t$ is plotted for four different starting states (at fixed $\mu_{s}=1.0$) subjected to the same applied shear, and all of which arrest at long times. However, the onset of arrest varies significantly between the four cases. In order to quantify a time for the onset of arrest, we define $t_{c}$ as the time at which the strain rate first drops below a minimal value of $10^{-9}$, as indicated by the horizontal dashed line in the figure. The trends presented hereafter are not sensitive to the choice of cutoff strain rate for determining granular arrest. Despite $t_{c}$ varying by more than an order of magnitude in Fig.~\ref{fig2}(c) (see dotted vertical lines, and a black arrow depicting the span of $t_{c}$), such variations are not observed in the evolution of density, which is invariant to the starting state, as seen in Fig.~\ref{fig2}(d).

Systematic investigation of the dependence of $t_{c}$ on friction and applied shear required significant computational effort. In order to get robust statistics, at least $10^{3}$ simulations were run for each value of friction and applied shear for at least $10^{7}$ time steps, requiring a total of $\sim\!10^{5}$ simulations. When the applied shear is significantly below the critical value, all systems arrested within the duration of the simulation. However, in the vicinity of the critical shear, some simulations did not arrest even when the applied shear was below the critical value. For such cases, additional systems were simulated to ensure at least $95$\% of the simulations resulted in the onset of arrest. The complementary cumulative distribution or survival function $C(t_{c})$ for different friction $\mu_{s}$ and stress ratio $\mu$ is shown in Fig.~\ref{fig3}, and indicates the fraction of simulations that arrested at time $t_{c}$ within the simulation run time. Significantly, the distribution of $t_{c}$ is heavy tailed and is well approximated by a log-normal probability density function for most values of $\mu$ and $\mu_{s}$, as determined by maximum likelihood estimation~\cite{alstott2014powerlaw}. Although a log-normal probability distribution function provides a better characterization of the underlying distribution as compared to exponential, power-law and stretched-exponential functions, we have not rigorously ruled out other distribution functions that can better describe the heavy-tailed data. The heavy tails of the distribution imply that the onset of arrest has a wide distribution. Moreover, the distribution of arrest times becomes wider as the flow-arrest transition is approached, thereby resulting in fewer arrested systems within a given simulation time. Therefore near the transition, a given granular system could potentially take an extremely long time to arrest even as the bulk stress state suggests that it almost certainly will arrest. An important implication of these results is that the flow-arrest transition in granular matter is highly unpredictable. 

 \begin{figure}[b!]
\includegraphics[width=3.2in]{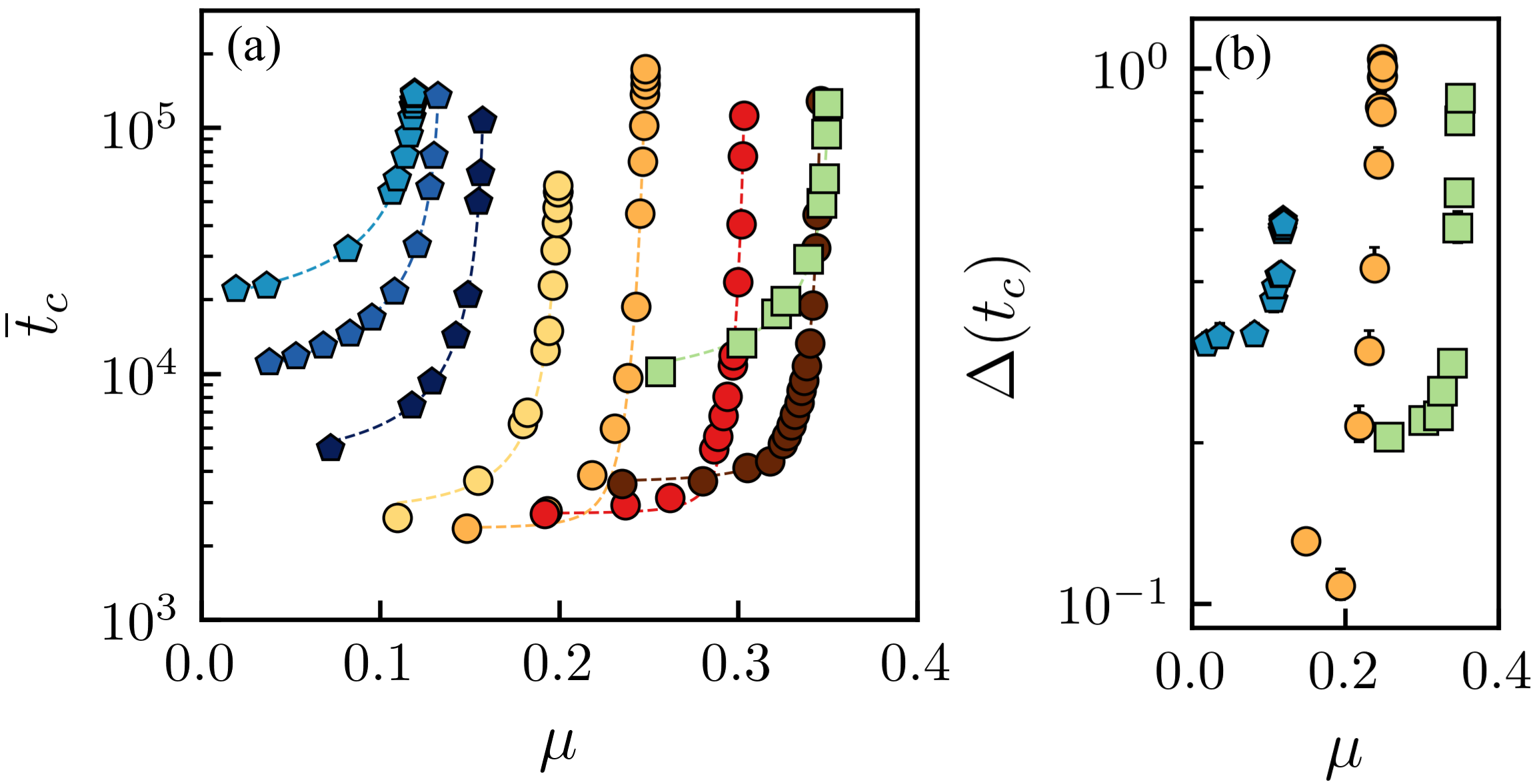}
\caption{(a) Variation of mean time to arrest $\overline{t}_{c}$ with $\mu$ for increasing values (left to right) of $\mu_{s}$ (see Fig.~\ref{fig5} for the value of $\mu_{s}$ corresponding to each symbol). The dashed lines correspond to the fit in Eq.~\ref{eq1} (b) Variation of the ratio $\Delta(t_{c})$ of standard deviation and mean time to arrest with $\mu$ for three different $\mu_{s}\!=\!10^{-3},6.5\!\times\!10^{-2},1.0$ from left to right. The error bars in (a) and (b) denote $95$\% confidence interval [error bars are smaller than the symbol in (a)].}
\label{fig4}
\end{figure}

\emph{Flow-arrest transition.---}We further characterize the distribution of arrest times by robust estimates of its moments---computed from $10^{5}$ bootstrapped samples of the simulation data~\cite{Efron:1986}---and associated confidence intervals of these estimates~\cite{efron1987better}. Good correspondence was observed between the estimated moments and those calculated from a best-fit log-normal distribution. The mean time to arrest $\bar{t}_{c}$ increases rapidly as the stress ratio $\mu$ is increased, and diverges at a friction-dependent critical value of shear for all values of friction, as shown on a semilog scale plot in Fig. 4(a). The rate of increase of $\bar{t}_{c}$ upon approaching $\mu_{c}$ also appears to be friction dependent, with small rates of increase for low and high friction, and larger rates of increase for intermediate friction. At low values of $\mu$ away from critical shear, the mean time to arrest saturates at a value that is indicative of the early-time transient evolution.

 \begin{figure}[b!]
\includegraphics[width=3.2in]{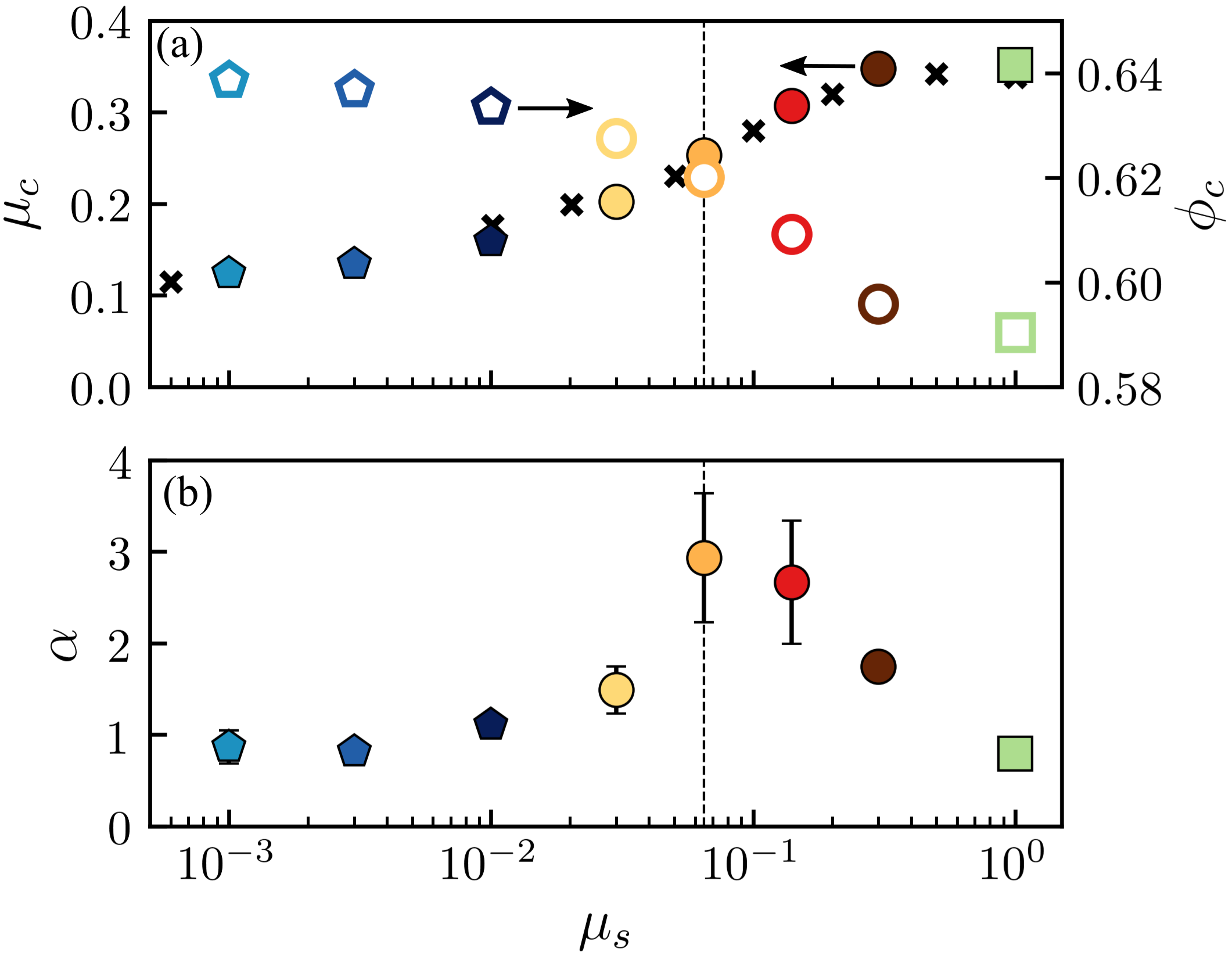}
\caption{(a) Critical stress ratio $\mu_{c}$ (closed symbols) and critical density $\phi_c$ (open symbols) as a function of friction $\mu_{s}$. Black crosses correspond to critical values of the stress ratio adapted from Singh \emph{et al}.~\cite{singh2013effect}. The leftmost black cross indicates the value for zero friction. (b) Variation of exponent $\alpha$ with friction $\mu_{s}$, as described in Eq.~\ref{eq1}. The error bars for closed symbols in (a) and (b) denote the standard deviation of power-law fitting error [error bars are smaller than the symbols in (a)]. The dashed line at the peak value of the exponent (bottom) coincides with the crossover values of $\mu_{c}$ and $\phi_{c}$ (top).}
\label{fig5} 
\end{figure}

Based on these observations, the variation of $\bar{t}_{c}$ with $\mu$ is well described by the following power-law form for all values of friction:
\begin{equation}
\overline{t}_{c} - t_{0} \sim \left(\mu_{c} - \mu \right)^{-\alpha},
\label{eq1}
\end{equation}
where $\mu_{c}$ denotes the critical stress at which mean arrest times diverge, and $\alpha$ is the exponent of the power law. $t_{0}$ is a fitting parameter that signifies the early-time transient evolution of the strain rate during which the density also evolves to its close-packed steady value as a consequence of starting from a dilute state, as shown in Fig.~\ref{fig2}. Although a similar power-law form of critical scaling below yield threshold was demonstrated in stress-controlled simulations of frictionless spheres~\cite{clark2018critical}, we do not rule out more complicated functional forms that could better represent the divergence of mean times, especially for intermediate frictions for which we observe deviations to Eq.~\ref{eq1} at small applied shear (see Fig.~S1 in Supplemental Material), and large fitting errors for the exponent as seen in Fig.~\ref{fig5}(b)~\footnote{The fitting of the data to Eq.~\ref{eq1} was performed using a Trust Region method for bounded non-linear optimization, followed by Levenberg-Marquardt non-linear least squares fitting.}.

Surprisingly, in addition to a diverging distribution mean, the second moment of the distributed data also diverges at the critical stress. In Fig.~\ref{fig4}(b) the ratio of standard deviation to mean $\Delta(t_{c})\!=\!\frac{\sqrt{\overline{t_{c}^{2}} - \overline{t}_{c}^{2}}}{\overline{t}_{c}}$ is plotted as a function of increasing shear stress. The diverging nature of $\Delta(t_{c})$ indicates that the distribution rapidly becomes wider and heavier-tailed upon approaching $\mu_{c}$, underscoring that in the vicinity of $\mu_{c}$ it is extremely difficult to predict when a flowing granular system will arrest. This observation has important implications for practical applications involving granular materials, such as clogging in granular flow through a hopper, which has been suggested to not exhibit a well-defined clogging transition~\cite{Thomas:2016bt}. 

The critical shear $\mu_{c}$ that defines the flow-arrest transition of granular matter is highly friction dependent. In Fig.~\ref{fig5}(a) $\mu_{c}$ is plotted as a function of $\mu_{s}$, and it is an increasing function of friction. Additionally, the friction-dependent critical value of $\mu_{c}$ is in excellent correspondence with critical yield stress calculated from flow-controlled simulations~\cite{singh2013effect}. The value of critical shear at high friction $\mu_{c}\!=\!0.35$ is also similar to the observed value of yield stress in experiments~\cite{Boyer:2011hf} and simulations~\cite{Salerno:2018en} on frictional granular matter. Corresponding to $\mu_c$, a friction-dependent critical density $\phi_c$ at the flow-arrest transition is also computed by averaging over densities for all arrested states at the largest simulated stress ratio $\mu$ below $\mu_c$. Figure~\ref{fig5}(a) demonstrates that $\phi_c$ monotonically decreases with increasing $\mu_s$. At low friction, $\phi_c\!=\!0.64$, and is similar to the random close packing density for monodisperse frictionless spheres. At high friction, $\phi_c\!=\!0.59$, and is similar to the critical density in critical state theory~\cite{schofield1968critical}, which has also been observed in experiments on frictional granular matter~\cite{Boyer:2011hf}.

Intriguingly, the exponent $\alpha$ of power-law divergence varies nonmonotonically with friction, as shown in Fig.~\ref{fig5}(b). At low and high friction, the exponent is close to unity, whereas its value is greater than one for intermediate friction. Additionally, this range of intermediate friction coincides with the crossover from the low-friction to high-friction regime, as seen by the variation of $\mu_{c}$ and $\phi_{c}$ with $\mu_{s}$ in Fig.~\ref{fig5}(a). Clark \emph{et al}., in their stress-controlled simulations observed a critical scaling near the yield transition of frictionless granular packings and extracted an exponent near unity~\cite{clark2018critical}. At present we do not have an explanation for the variation of the exponent. The exponent is sensitive to the statistics of arrest times, and large errors were observed in its fitting for intermediate friction, as shown by the errors bars in the figure. Additionally, the arrest time statistics are likely sensitive to system size, and finite size effects constitute an important area of focus for future work. 

Although the foregoing technical issues require further analysis, recent theoretical work has indeed predicted three regimes of frictional granular rheology with a nonmonotonic dependence of kinetics on friction, especially correlated velocity fluctuations of the particles~\cite{DeGiuli:2016ew}.  However, the theory is formulated for steady granular flows. Extensions of the model to transient nonequilibrium flows can determine whether the nonmonotonic variations observed here result from friction-dependent kinetics. 

The statistical nature of flow-arrest granular transition described here is not captured by current deterministic models that predict equilibrium granular plasticity and rheology. The wide distribution of times to arrest near the transition are apparently sensitive to microscopic details, such as stress and strain rate heterogeneity within the microstructure, and these details need to be incorporated for accurately predicting the transition. Previous simulations on 2D frictional granular matter demonstrated mechanical heterogeneity during transition from quasistatic to inertial flows, along with a diverging correlation length~\cite{Gimbert:2013js}. Although we have demonstrated diverging times during transient flow-arrest transition, it remains to be tested whether a diverging length scale also exists. Additionally, the approach to arrest transition from a flowing state is accompanied by a massive drop in strain rate, which continues its decay with time. At small strain rates in steady granular flows, local plastic events are spatially correlated~\cite{LeBouil:2014hp}, and this is often interpreted by a nonlocal formulation of granular rheology~\cite{Kamrin:2012du}. Extensions of such steady-state nonlocal rheological formulations to include transient effects~\cite{Henann:2014cp} could provide a meaningful way to interpret the present results. 

This work was performed at the Center for Integrated Nanotechnologies, a U.S. Department of Energy and Office of Basic Energy Sciences user facility. Sandia National Laboratories is a multimission laboratory managed and operated by National Technology and Engineering Solutions of Sandia, LLC, a wholly owned subsidiary of Honeywell International, Inc., for the U.S. Department of Energy's National Nuclear Security Administration under Contract DE-NA-0003525.  
\bibliographystyle{apsrev4-1}

\end{document}